\documentclass[superscriptaddress,aps,pre,twocolumn]{revtex4-1}
\usepackage[utf8]{inputenc}
\usepackage{amsmath}
\usepackage{amssymb}
\usepackage{amsthm}
\usepackage{thmtools}
\usepackage{refcount}
\usepackage{siunitx}
\usepackage{graphicx}
\usepackage{hyperref}
\hypersetup{colorlinks,allcolors=black}
\usepackage{xcolor}

\usepackage{natbib}
 \usepackage[capitalise]{cleveref}
\usepackage{microtype}
\usepackage{bm}
\usepackage{lipsum}
\usepackage[english]{babel}
\usepackage{color}
\usepackage{graphicx}
\usepackage{xcolor}
\usepackage{braket}

\newcommand{\bp}{\mathbf{p}}

\begin{document}
\author{Haidar Al-Naseri}
\email{hnaseri@stanford.edu}
\affiliation{Stanford PULSE Institute, SLAC National Accelerator Laboratory, Menlo Park, California 94025, USA}

\author{Gert Brodin}
\email{gert.brodin@umu.se}
\affiliation{Department of Physics, Ume{\aa} University, SE--901 87 Ume{\aa}, Sweden}
\title{Applicability of semi-classical theories in the strong field plasma regime}

\begin{abstract}
For many purposes, classical plasma dynamics models can work surprisingly well even for strong electromagnetic fields, approaching the Schwinger critical fields, and high frequencies, approaching the Compton frequency. However, the applicability of classical models tends to depend rather sensitively on the details of the problem.  In the present paper, we study the specific case of plasma oscillations to draw a line between the classical and quantum relativistic regimes. Due to the field geometry of study, mechanisms like radiation reaction and Breit-Wheeler pair production, which tend to be important for electromagnetic fields, are rather effectively suppressed. Moreover, we find that the polarization current due to the electron spin is generally negligible for frequencies below the Compton frequency, compared to the free current, whose magnitude is well-approximated by the classical Vlasov theory. However, we show that pair creation due to the Schwinger mechanism can sometimes be important for surprisingly modest field strengths, of the order of 10 \% of the critical field or even smaller. A rough guideline for when the classical Vlasov theory can be applied is given.    
   
\end{abstract}
 
\maketitle
\section{Introduction}

The interest in strong field physics in the plasma and QED communities has increased in the past decades due to the rapid growth and development of various laser facilities \cite{fedotov2023advances,DiPiazza}. This large research activity has resulted in theoretical and experimental success. The latter includes e.g. observation of the pair creation in SLAC \cite{SLAC1997} and the observation of quantum radiation reaction in Rutherford \cite{Rutherford}. There are also large efforts to probe the non-perturbative strong field QED regime using the collision of an intense laser field with a multi GeV electron beam in E320 in SLAC \cite{SLAC} and the LUXE project in Desy \cite{LUXE}. 

Despite the steady progress of laser intensities in the current experimental facilities, the peak electric field for the current laser in the lab frame is still well below the Schwinger limit $E_{cr}=m^2 c^3/e\hbar \sim 1.3 \times 10^{18} V/m$ \cite{Schwinger}, where $m$ and $e$ are the electron mass and charge, respectively. However, it can be noted that there are proposals to use plasma-based schemes to enhance the peak field to approach the Schwinger limit \cite{Vinventi2021,Vincenti2024}.

A common computational method to simulate strong field QED effects is using particle-in-cell (PIC) codes, which at the core account for the Maxwell-Vlasov dynamics, but with various extensions (as, for example, in the codes of Refs. \cite{OSIRIS,PICADOR}) featuring modules covering additional strong field physics.  Such modules include e.g. radiation reaction (in the classical or quantum version), and nonlinear Breit-Wheeler pair creation. These codes have been used to simulate the collision of high-intensity lasers with solid targets \cite{Laser-solid,Vincenti2024}, relativistic electron beams \cite{Electro-beam}, and purely optical setups \cite{Only-Optical1,Only-Optical2}. In particular, in recent works \cite{Pinching, Vincenti2024}, electromagnetic fields with peak electric fields up to $0.5 E_{cr}$ have been used in strong field PIC simulations. A natural question, in this context, is to what degree the underlying classical framework, even if modified, can be applied when the most extreme fields are approached.  

The breakdown of classical Vlasov theory due to strong radiation reaction is quite well understood (see e.g. \cite{blackburn2020radiation,burton2014aspects,dinu2016quantum}). Similarly, pair-creation effects due to nonlinear Breit-Wheeler processes have received much attention \cite{fedotov2023advances,di2016nonlinear,seipt2020spin}. In the present work, we will apply the Dirac-Heisenberg-Wigner (DHW) formalism \cite{Birula,2021} to study strong field effects in an electrostatic field geometry, when both these processes are suppressed to a good approximation.  Strong field effects that still will be present include pair-creation due to the Schwinger mechanism \cite{Schwinger} (or the dynamically assisted Schwinger mechanism \cite{schutzhold2008dynamically}), vacuum polarization currents \cite{king2016measuring,brodin2003light}, and spin polarization effects \cite{crouseilles2023vlasov,hussain2014weakly}. To compare our results with the classical case, we use the relativistic Vlasov to model the plasma dynamics.
As is well-known, the Vlasov theory breaks down for fields close to the critical field, due to the pair-creation by the Schwinger mechanism. However, depending on the plasma density, there can also be a breakdown associated with Schwinger pair creation for much smaller fields, of the order $10\%$ of the critical field, or even smaller. Breakdown of the Vlasov theory for comparatively weak fields requires that the plasma density is small (or at least modest) initially, such that a relatively small number of created pairs can make a difference for the plasma dynamics. In principle, there can also be a breakdown of the Vlasov theory due to other effects, e.g. spin polarization, but it is shown that this will require parameters of a more extreme nature.  

The organization of the paper is as follows:
In \cref{DHW-formalism}, we briefly describe the DHW-formalism and present the equation system to be solved numerically. Preliminaries regarding the classical Vlasov treatment are presented in \cref{Vlasov_section}. Next, in \cref{Numer_sections}, the results from the numerical comparison are shown. Finally, in \cref{Discussion} we make a summary of the paper together and present the conclusions.

\section{Basic equations}

Here we outline the basic classical and quantum models to be analyzed and compared in the rest of the work.
\subsection{The DHW-formalism}
\label{DHW-formalism}

The DHW-formalism is a quantum kinetic description of the Dirac field that was first presented in Ref. \cite{Birula}. It is based on a Gauge invariant Wigner transformation of the density matrix built from the Dirac four spinors. The main assumption used in the DHW-formalism is the Hartree-Fock (mean-field) approximation, which is applicable for the electromagnetic field. Besides the presentation made in the original paper \cite{Birula}, slightly different derivations have been given in Refs.
\cite{Gies,Sheng}. Moreover, a shortened summary of the main steps of the derivation has been given in, for example, \cite{2021}.

The full set of the DHW equations consists of 16 coupled scalar phase-space variables, that describe mass (phase-space) density, charge density, current density, spin density, etc. However, for the specific case of a simplified electrostatic 1D geometry, with the electric field $\mathbf{E}=E(z,t)\mathbf{\hat{z}}$, the
full system (see Ref. \cite{2021}) can be reduced to 4 coupled scalars according to

\begin{align}
\label{PDE_System}
    D_t\chi_1(z,\bp,t)&= 2\varepsilon_{\bot} \chi_3(z,\bp,t)- \frac{\partial \chi_4}{\partial z} (z,\bp,t)\notag\\
    D_t\chi_2(z,\bp,t) &= -2p_z\chi_3(z,\bp,t)\\
    D_t\chi_3(z,\bp,t)&= -2\varepsilon_{\bot}(p_{\bot}) \chi_1(z,\bp,t) +2p_z\chi_2(z,\bp,t)\notag\\
    D_t\chi_4(z,\bp,t)&= -\frac{\partial \chi_1}{\partial z}(z,\bp,t)\notag 
\end{align}
together with the Ampère's law
\begin{equation}
\label{Ampers_law}
\frac{\partial E}{\partial t}=-\frac{e}{(2\pi)^3}\int \chi_1 d^3p
\end{equation}
where $D_t=\partial/\partial t +e{\tilde E}\partial/\partial p_z$ and $\varepsilon_{\perp}=\sqrt{m^2+p_{\perp}^2}$, and where $p_{\perp}$ is the perpendicular momentum and ${\tilde E}=\int_{-1/2}^{1/2} E(z+i\lambda\partial/\partial p_z,t)d\lambda$. The $\chi_i$ (i=1,2,3,4) are dimensionless variables that have the following physical interpretation: Firstly, $\chi_1$ is the current density

\begin{equation}
    j_z=\frac{e}{(2\pi)^3}\int \chi_{1} d^3p
\end{equation}
where $j_z$ is the current.
The second variable $\chi_{2}$ gives the mass density $\rho_m$, i.e.
\begin{equation}
    \rho_m =\frac{m}{(2\pi)^3}\int \frac{\chi_{2}}{\varepsilon_{\perp} } d^3p
\end{equation} 
Furthermore, $\chi_{3}$ gives the spin density, i.e. the angular momentum density ${\bf M}$ due to the spin is 
\begin{equation}
    {\bf M} =\frac{1}{(2\pi)^3}\int ({\bf \hat{z}}\times {\bf p}) \frac{\chi_{3}}{2\varepsilon_{\perp} } d^3p
\end{equation} 
and finally $\chi_{4}$ gives the charge density $\rho_c$, i.e. 
\begin{equation}
    \rho_c=\frac{e}{(2\pi)^3}\int \chi_{4} d^3p.
\end{equation}
Note that for spatial scale lengths much longer than the characteristic de Broglie length (and naturally, also for the spatially homogeneous case), the nonlocal electric field $\tilde{E}$ simply reduces to the ordinary electric field $E(z,t)$. The system of equations has been presented in natural units with $\hbar=c=1$.   

Next, we rewrite the system \cref{PDE_System} to
 simplify the numerical calculations. In this context it is important to note that $\chi_1$ and $\chi_2$ have nonzero vacuum contributions, associated with the expectation values of the free Dirac field operators, (see e.g. Ref. \cite{Birula,plasma_quantum_kinetic_review}) where the vacuum expressions are given by
  \begin{align}
 \chi_{1vac}&=-\frac{2 p_z}{\varepsilon}\notag \\ \chi_{2vac}&=-\frac{2\varepsilon_{\perp}}{\varepsilon}.
 \end{align}
 Here $\varepsilon=\sqrt{m^2+p^2}$, and we introduce new variables $\Tilde{\chi}_i(z,\bp,t)$ as the deviation from the vacuum state, i.e. we let
 \begin{equation}
     \Tilde{\chi}_i(z,\bp,t)=\chi_i(z,\bp,t)- \chi_{i {\rm vac}}(\bp).
    \end{equation}
    Note that $\Tilde{\chi}_{3,4}={\chi}_{3,4}$. 
 Secondly, we switch to the canonical momentum. Using the Weyl gauge where the scalar potential is zero such that $E=-\partial A/\partial t$, the $z$-component of the kinetic momentum is replaced by $q=p_z+eA$. With $q$ as one of the independent variables, the operator $D_t$ simplifies to $D_t=\partial/\partial t$. Our third modification is to switch to dimensionless variables.  The normalized variables are given by, $t_n= \omega_c t$, $q_n=q/mc$, $p_{n\perp}=p_{\perp}/mc$, $E_n=E/E_{cr}$, $A_n=eA/mc$, where $\omega_c =(mc^2/\hbar)$ is the Compton frequency, and we note that the DHW-functions are already normalized. For notational convenience, we omit the index $n$ and $\Tilde{}$ in what follows. Finally, we consider the homogeneous limit,  making the variable $\chi_{4}=0$. The equations to be solved numerically now read: 
\begin{align}
    \frac{\partial  \chi_1}{\partial t} (q,p_{\bot},t)&= 2\varepsilon_{\bot} \chi_3 + 2E\frac{\varepsilon_{\bot}^2}{\varepsilon^3} \notag\\
      \frac{\partial  \chi_2}{\partial t} (q,p_{\bot},t )&= -2(q-A)\chi_3-2(q-A)E\frac{\varepsilon_{\bot}}{\varepsilon^3}\notag\\
      \frac{\partial \chi_3}{\partial t} (q,p_{\bot},t)  &=
      -2\varepsilon_{\bot} \chi_1 
      +2(q-A)\chi_2 
     \label{system}
      \end{align}
      with Ampère's law
      \begin{equation}
\label{Ampers_law2}
\frac{\partial E}{\partial t}=- \eta \int \chi_1 d^2p
\end{equation}
where we hade defined the dimensionless factor $\eta=\alpha/\pi\approx 2.322 \times10^{-3}$, where $\alpha$ is the fine-structure constant. Note that the ions are considered to be a neutralizing background, and we do not consider their motion.
Due to cylindrical symmetry, the azimuthal integration has already been carried out, and hence $d^2p=p_{\bot}dqdp_{\bot}$.

Before ending this subsection, let us briefly discuss what physics that is included in the coupled system of \cref{system} and \cref{Ampers_law2}. Due to the mean-field approximation, as expected, single-particle Larmor emission will not be contained in the model, nor will processes involving few particles (of the Dirac-field) such as Breit-Wheeler pair-production be covered.
The choice of the 1D electrostatic field geometry makes it possible to have both Breit-Wheeler pair-production and the radiation reaction negligible compared to the collective plasma effects. For more details, including calculation justifying the above statement, see Ref.\cite{2023}.

We note that by dropping all quantum effects, the model reduces to the 1D electrostatic limit of the relativistic Vlasov equations. However, all quantum effects that can be described as a a collective phenomenon are generally included in the model. Specifically this include e.g. collective pair-creation, collective pair-annihilation, Pauli-blocking, spin-polarization and vacuum effects such as a finite vacuum polarization \cite{linear}. The vacuum contribution also gives rise to the issue of charge renormalization. However, this turns out to be a negligible effect when ruining a numerical simulation with a cutoff in the momentum-space, for more details, see \cite{2023}.

\subsection{The Vlasov theory}
\label{Vlasov_section}
In this subsection, we present the relativistic Vlasov theory that will be used throughout the work as a classical comparison with the DHW model. We use the same normalization for the relativistic Vlasov equation as in the DHW theory and naturally, we apply the same field geometry as in the previous subsection.  For the 1D electrostatic fields, the electron distribution function $ f_e$ obeys: 
\begin{equation}
\label{CircularPol}
    \frac{\partial f_e}{\partial t}
    +  eE_z \frac{\partial f_e}{\partial p_z}=0
\end{equation}
This system is closed by Ampere's law in the form
\begin{equation}
\label{Ampers}
    \partial_tE_z=-4\pi e \int d^3p \frac{p_z}{\epsilon}f_e
\end{equation}
Note that the ions are considered to be as a neutralizing background and we do not consider their motion. Again the numerical solution is simplified by using the canonical transformation (using normalized units),
\begin{align}
q&=p_{z} +A \label{canon}
\end{align}
Using the canonical transformation in \cref{canon}, we get
\begin{equation}
\label{Vlasov_canon.}
    \frac{\partial f_e }{\partial t}(p_{\perp},q,t)=0
\end{equation}
Finally, the Vlasov equation is coupled to Ampere's law \cref{Ampers}
\begin{align}
\label{Amper_norm}
    \frac{\partial E}{\partial t}&=-2\eta \int d^2p \frac{q-A }{\epsilon}f_e
\end{align}
which gives us a closed system when expressing the electric field in terms of $A$. Given the initial values of the numerical setup for any distribution function that is initially $f_e=F(p_{\perp},q)$, the time-dependent solution to \cref{Vlasov_canon.} is $F(p_{\perp},q-A(t))$. Here, $A$ is the sum of external and self-consistent vector potential. That is to say, we update the initial distribution function by shifting the $q$ dependence with $A(t)$. Furthermore, we have to solve Ampere's law for each time step. 

\section{Numerical results}


\label{Numer_sections}
\subsection{Preliminaries}

Both the DHW-equations, \cref{system} with the Ampere's law \cref{Ampers_law2}, and the classical system \cref{Vlasov_canon.,Amper_norm} have been solved numerically using a phase-corrected staggered leapfrog method \cite{LeapFrog}. We have considered three independent variables, time, longitudinal and perpendicular momentum ${t,q,p_{\perp}}$, where we have used typical parameters of time step $\Delta t=0.002$, $\Delta q=0.01$, and $\Delta p_{\perp}=0.1$. Even though we are solving a numerical problem that is homogeneous,
a typical run with strongly relativistic motion, requiring a large cutoff $ q_{max}> \gamma_{max}$, where $\gamma_{max}$ is the maximum relativistic gamma factor of the particles, is memory-demanding.
The soundness of the numerical solution has been tested by studying the energy conservation, based on the conservation laws, 
\begin{align}
    &\frac{d}{dt}\bigg( \frac{E^2}{2}+ \eta\int d^2p \Big[\chi_2+ (q-A)\chi_1 \Big] \bigg)=0\\
    & \frac{d}{dt} \bigg(\frac{E^2}{2}+ 2\eta \int d^2p \sqrt{1+p_{\perp}^2+(q-A)^2}f_e(q,p_{\perp}) \bigg)=0
\end{align}
for the DHW system and the Vlasov systems, respectively. Both of the conservation laws are fulfilled to a good approximation in the simulations, with relative numerical errors typically less than $10^{-4}$.

As an initial condition for the plasma, we use a Fermi-Dirac distribution $f_{FE}$
\begin{equation}
    f_{FE}=\frac{1}{1+e^{(\epsilon-\mu_n)}/T_n}
\end{equation}
where $\mu_n$ and $T_n$ are the normalized chemical potential and temperature. This distribution can be directly applied in the Vlasov case, that is we set $f_e(t=0)=f_{FE}$. However, for the DHW theory, we must use 
\begin{align}
    \chi_1&=\frac{2 q f_{FE}}{\sqrt{1+q^2+p_{\perp}^2}}\\
    \chi_2&=\frac{2 \sqrt{1+p_{\perp}^2} f_{FE}}{\sqrt{1+q^2+p_{\perp}^2}}\\
    \chi_3&=0
\end{align}
where the last relation corresponds to no spin polarization in the initial distribution. 
For the initial condition of the electric field field, in principle, we could start with the field amplitude $E_0$ as $E(t=0)=E_0$. However, in the DHW theory, this will yield a problem (whose magnitude can be more or less severe, depending on the parameters) as in general, we will get pair creation in the first time step, due to a sudden change from vacuum to a maximum value of the field. Thus, in order to avoid this, we introduce the field smoothly into the system by adding an external current source $J_{ex}$
\[J_{ex}= \begin{cases} 
      \frac{E_0}{T} &0< t\leq T \\
     0& T<t \\
   \end{cases}
\]
that is added as a source in Ampere's law. This current will introduce a field that reaches its maximum values at $t=T$. After the initial setup ending at $t=T$, we will proceed with a simulation with an effective starting value $E(t=T)=E_0$. Thus we study the same physics as if starting with $E(t=0)=E_0$, but we avoid the initial time step that can be potentially problematic.

\subsection{The polarization current}
We start our numerical investigation by looking at the details of the polarization current density, which has no correspondence in classical theory.
The polarization current in DHW theory is given by
\begin{equation}
j_p=\frac{\partial}{\partial t} \int\frac{  \chi_3}{2\epsilon_{\perp}} d^2p 
\end{equation}
The polarization current contains both the vacuum polarization current as well as the particle polarization current, where the latter arises due to the spin polarization of the particles. It is easy to show that the standard expressions for the vacuum polarization currents (see e.g. Ref. \cite{marklund2006nonlinear}) will not be significant for fields strengths $E\ll E_{cr}$ and for frequencies below the Compton frequency. When these conditions are violated, we will have significant pair production. Thus, from now on, we will assume that the vacuum polarization current will be negligible compared to the free current. Nevertheless, the vacuum contribution is of theoretical interest, as it leads to UV divergencies in the momentum integral for the current, which in principle must be fixed through a renormalization.  We will not go through the renormalization here, as it has little practical consequences in a numerical calculation. Due to the finite momentum cut-off acting as a regularization, and the slow logarithmic divergence, the contribution from a renormalization will only affect the numerical coefficient in Ampere's law marginally, see Ref. \cite{2023} for further details. However, since we would like to address the particle spin-polarization current, it will be illustrative to do some analytical calculations to separate the particle contribution to the polarization current from the vacuum contribution. 

Following Ref. \cite{2023}, we can combine the three equations in the system (\ref{system}) and solve for $\chi_3$, treating the operator $(1/\varepsilon )\partial _{t}$ as a small expansion parameter, which is valid for frequencies below the Compton frequency. This results in the expression
$\chi_3$ for $\frac{\chi_{3}}{2\epsilon_{\perp}}\,\ $we get%
\begin{equation}
\chi_{3}=-D^{-1}\left[ \frac{4\varepsilon _{\bot }E}{\varepsilon }-2E%
\chi_{2}\right]   
\label{chi3-solved}
\end{equation}
where the inverse operator is 
\begin{equation}
D^{-1}=\left[1-\frac{1}{4\varepsilon ^{2}}\frac{\partial ^{2}}{\partial
t^{2}}+ \left( \frac{1}{4\varepsilon ^{2}}\frac{\partial ^{2}}{\partial t^{2}} \right)^2%
+...\right] \frac{1}{4\varepsilon ^{2}}  \label{InverseD}
\end{equation}%
where we have kept up to fourth-order corrections, but this can easily be done to all orders. In Eq. (\ref{chi3-solved}) it is straightforward to separate the vacuum polarization (the first time inside the bracket) from the particle polarization (the second term inside the bracket proportional to the mass density of real particles $\chi_2$). Furthermore, for frequencies well below the Compton frequency, it is enough to keep the lowest order term of Eq. (\ref{InverseD}). Given this, we can define the lowest-order particle polarization current $j_{pp0}$ as
\begin{equation}
j_{pp0}=\frac{\partial}{\partial t} \int  \frac{\chi_2 E}{2\epsilon_{\perp}\epsilon^2} d^2p 
   \label{lowest}
\end{equation}
which is the current (in the leading low-frequency approximation) due to the spin-polarization of particles.

\begin{widetext}

\begin{figure}
    \centering
    \includegraphics[width=\textwidth]{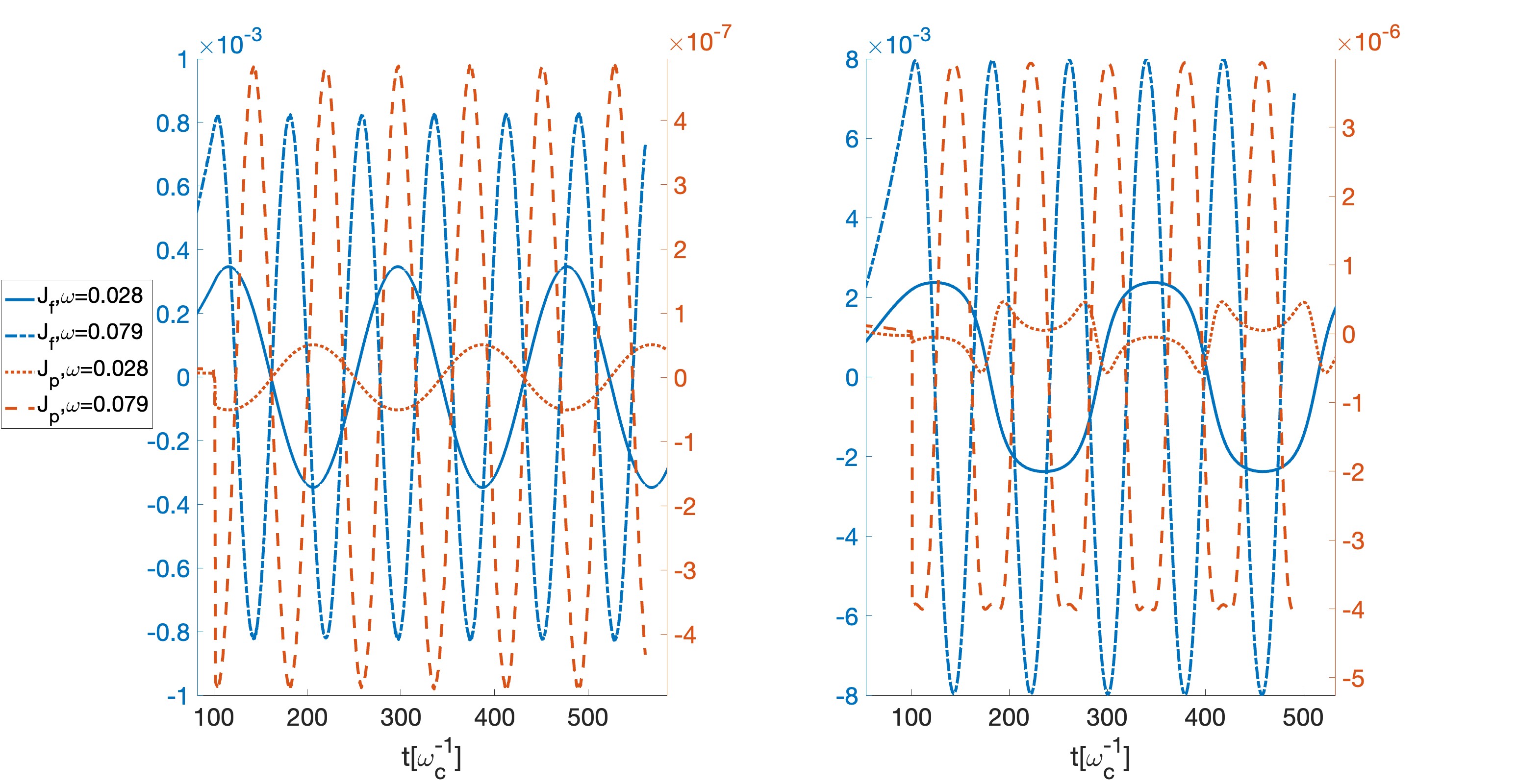}
    \caption{The free- and polarization- currents are plotted versus time. We have used plasma densities corresponding to plasma oscillation frequencies $\omega=0.028 \omega_c$ and $\omega =0.079\omega_c$. In the left panel, we used $E=0.01 E_{cr}$ and in the right panel we had $E=0.1E_{cr}$}
    \label{Fig1}
\end{figure}
\end{widetext}

Solving the DHW equations, we can compare the free part of the current density
\begin{equation}
    j_{f}=\int \frac{q-A}{\epsilon_{\bot}}\chi_2d^2p
\end{equation}
with the part due to the particle spin polarization $j_{pp0}$.  In Fig \ref{Fig1} we have plotted the time dependence of the free current density and the particle spin polarization current density for two different plasma densities, leading to different plasma oscillation frequencies. Note that the scale to the right applies for $j_{pp0}$, which in both cases is much smaller than the free current density. In the left panel, we have used a more modest electric field amplitude, $0.01E_{cr}$, resulting in an almost harmonic temporal profile for all currents. The main physical feature to note is the different frequency dependencies of the two contributions, where the relative contribution of the particle polarization current increases with frequency. However, even for frequencies approaching the Compton frequencies, $j_{pp0} \ll j_{f}$ would still apply. In the right panel, we have considered a substantially stronger electric field amplitude, $0.1 E_{cr}$. For the lower frequency in the right panel, it is clear that this leads to a non-harmonic temporal dependence. However, the frequency scaling is similar to the case of weaker fields, i.e. the relative magnitude of the particle polarization current $j_{pp0}$ increases with frequency. Still, the conclusion that the particle polarization current is small compared to the free current, applies also for stronger fields, as long as $E<E_{cr}$.    

\begin{figure}
    \centering
    \includegraphics[width=9 cm, height=8 cm]{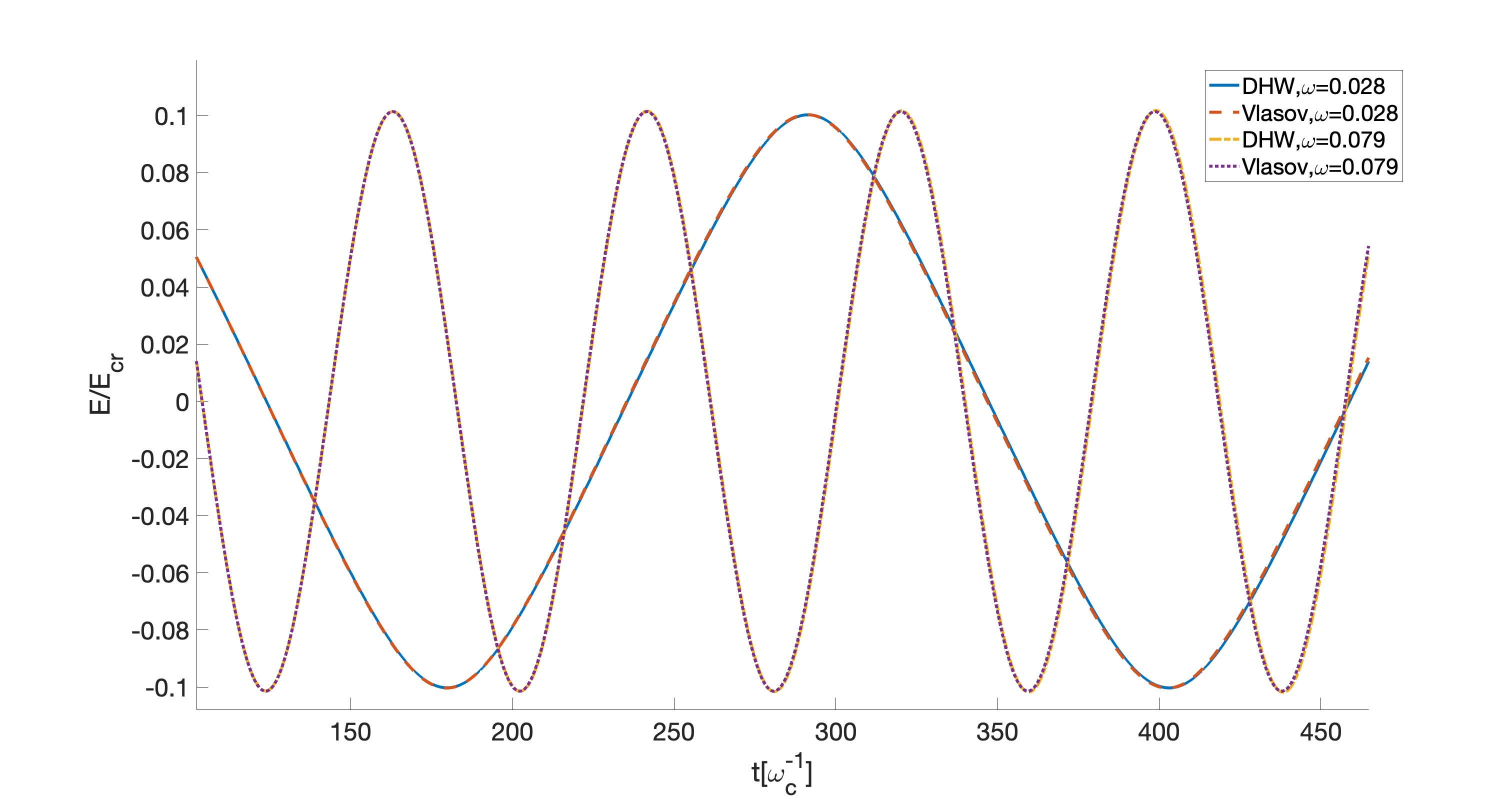}
    \caption{The electric field $E/E_{cr}$ is plotted over time for the DHW and Vlasov models, using an initial field amplitude $E_0=0.1 E_{cr}$ and plasma frequencies $\omega=0.028\omega_c$ and $\omega=0.079 \omega_c$.}
    \label{Fig2}
\end{figure}

While the omission of the polarization current from particles compared to the free current has broad applicability, this does not necessarily mean that the classical Vlasov equation is a good approximation. However, solving the Vlasov and DHW  equations for the same parameters as in  \cref{Fig1}, the evolutions of the self-consistent field in the Vlasov and the DHW models coincide very well, see \cref{Fig2}, where the Vlasov and DHW curves more or less coincide and the deviation is hardly visible. Since the plasma frequency is well below the Compton frequency, and the electric field amplitude is only $0.1 E_{cr}$, such that Schwinger pair creation is exponentially suppressed, one might think that this is sufficient conditions for the Vlasov model to be accurate. As we will see in the next subsection, however, a further condition is needed to assure the applicability of the Vlasov approximation.  

\subsection{Pair production}

Pair production due to the Schwinger mechanism is a well-known phenomenon, which, of course, is outside of classical descriptions. However, since this process is exponentially suppressed for fields below the critical field \cite{Schwinger}, at first one may think that this particular mechanism cannot be important for electric fields well below the critical field. It turns out that this is not quite true. 

In this subsection, we will show that a significant departure from the classical theory is possible even for comparatively weak fields. For this purpose, we focus on the effects of pair productions in the plasma oscillation dynamics. While classical models lack the physics of pair creation from vacuum, it is still important to know for which parameters classical models are valid to describe the plasma dynamics, and thus we will compare the evolutions using the DHW and Vlasov models.    

\begin{figure}
   \centering
    \includegraphics[width=9cm, height=10cm]{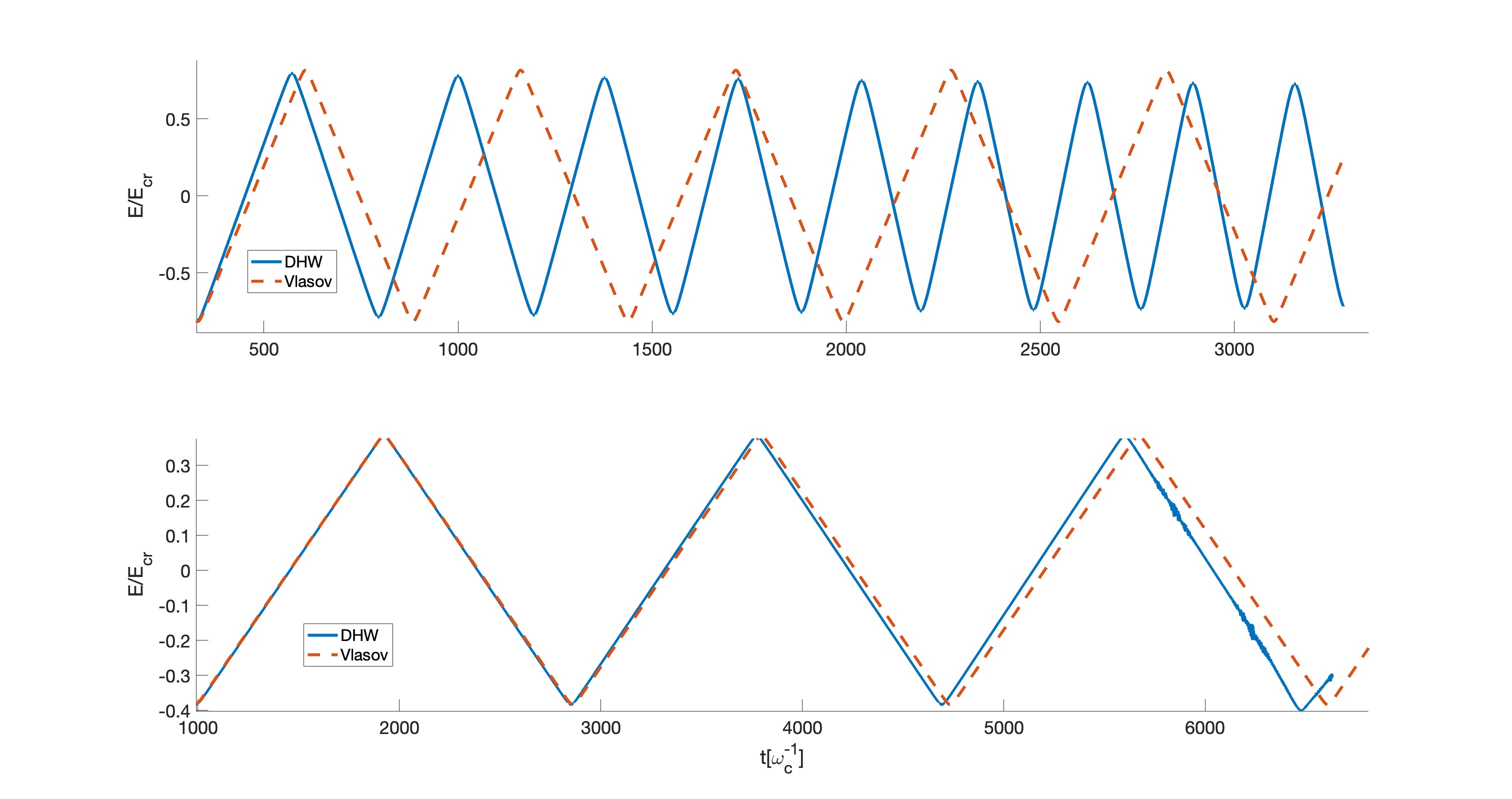}
    \caption{The electric field $E/E_{cr}$ is plotted over time from DHW and Vlasov. In the upper panel we have used $E_0=0.82 E_{cr}$ and $n=5.8\times 10^{29}/{\rm cm^3}$  while in the lower panel we used $E_0=0.38 E_{cr}$  and $n=8\times 10^{28}/{\rm cm^3}$.   }
   \label{Fig3}
\end{figure}
We start by comparing plasma oscillations using the DHW and Vlasov models for slightly stronger fields, where we do not expect classical theory to be a particularly good approximation. In \cref{Fig3}, we plot the electric field of the plasma oscillation as a function of time for two initial field amplitudes. In the first panel we have used $E(t=0)=0.82 E_{cr}$ and a plasma density $n=5.8\times 10^{29}/{\rm cm^3}$ while in the lower panel we had $E=0.38 E_{cr}$ and a plasma density $n=8\times 10^{28}/{\rm cm^3}$. For the upper panel, it is clear that the quantum model gives a gradually increasing frequency compared to the classical model, due to the electron-positron pairs being created as expected from the Schwinger mechanism. The difference is not pronounced after a single oscillation, but the period time in the quantum model decreases over time as the number of particles in the system increases. One can also see that the peak values of the electric field in the quantum model decrease slightly over time for the case when $E(t=0)=0.82 E_{cr}$. This is because the field energy is reduced over time when new electron-positron pairs are created.
In the lower panel for $E=0.38 E_{cr}$, the difference between the quantum model and the classical one is less clear. However, the same basic behavior is present, although the small decrease in electric field amplitude of the DHW-computation is too small to be visible in the lower panel of Fig. \ref{Fig3}. 

\begin{figure}
    \centering
    \includegraphics[width=10cm, height=10cm]{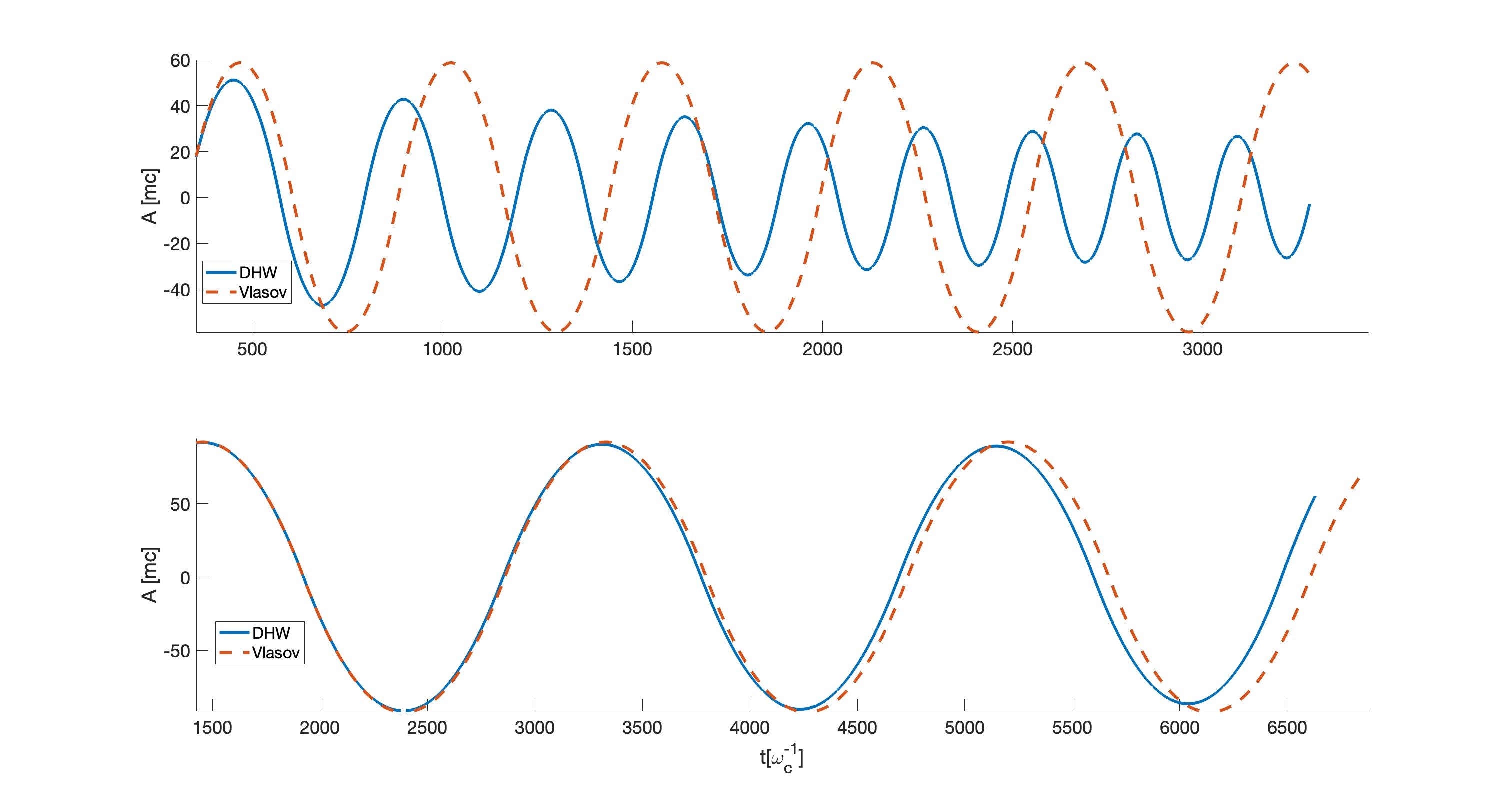}
    \caption{The Vector potential $A$ is plotted over time for the DHW and Vlasov models. In the upper panel we have used $E_0=0.82 E_{cr}$ and $n=5.8\times 10^{29}/{\rm cm^3}$  while in the lower panel we used $E_0=0.38 E_{cr}$  and $n=8\times 10^{28}/{\rm cm^3}$.  }
    \label{Fig4}
\end{figure}
 The deviation between the classical and quantum theory is more pronounced when studying the vector potential. In  \cref{Fig4} we plot the vector potential $A$ as a function of time using the same inputs as in \cref{Fig3}. Given the field geometry, the vector potential $A$ (besides giving the electric field) can be considered as the average momentum of the particles in the direction along the electric field. In the first panel of \cref{Fig4}, $A$ in the quantum model is gradually reduced compared to the classical result with no damping. This is because the increase in the number of particles in the system gives a higher plasma frequency, which in turn reduces the time the particles are being accelerated in the positive or negative direction. Hence, the particles reach a lower maximum momentum (maximum $A$) compared to the classical model. For the lower panel, we have a modest deviation of the average momentum only. Still, although the effect is much less pronounced for the lower value of the electric field, it is clear that the quantum model predicts a deviation that is gradually increasing with time as the number of pairs created accumulates.

\begin{figure}
    \centering
    \includegraphics[width=9cm, height=9cm]{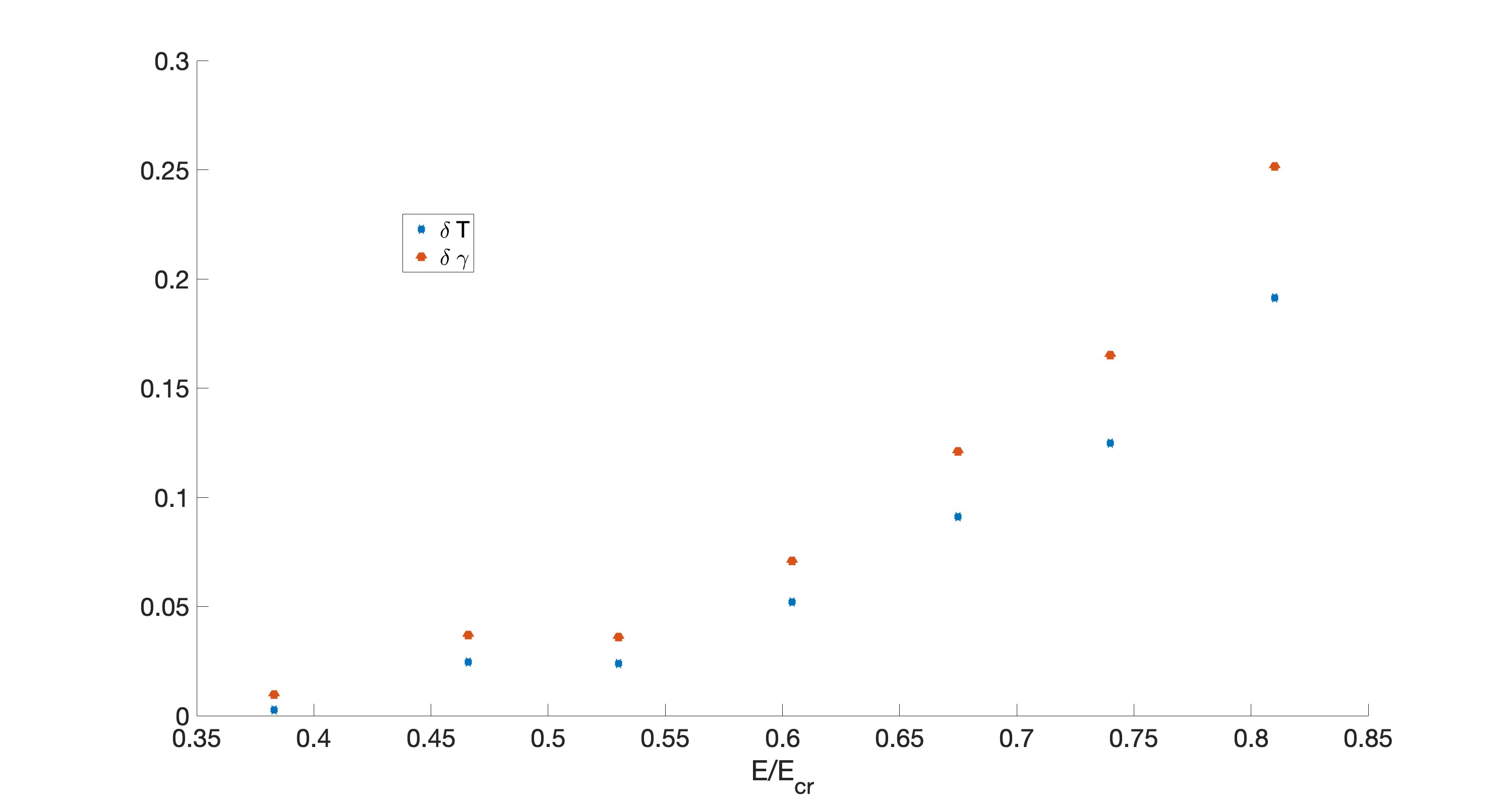}
    \caption{The relative errors in period time $\delta T$  and average gamma factor $\delta \gamma$ is plotted versus the electric field $E/E_{cr}$ for an initial plasma with the temperature $T=2.5$ and the plasma density $n=2\times 10^{29}/{\rm cm^3}$.  }
    \label{Fig5}
\end{figure}

\begin{figure}
    \centering
    \includegraphics[width=9cm, height=9cm]{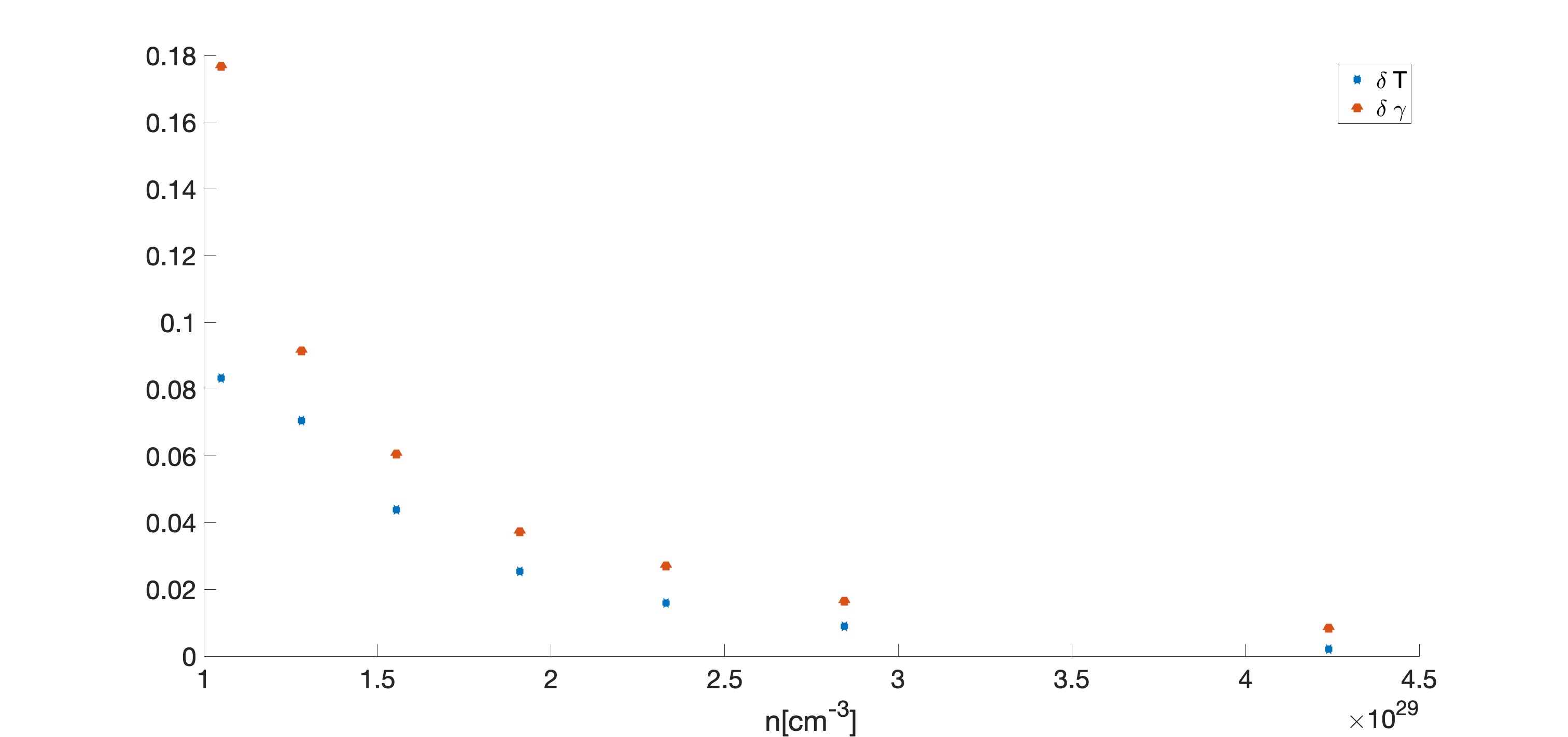}
    \caption{The relative errors in period time $\delta T$ and average gamma factor $\delta \gamma$ is plotted versus plasma density $n$ for an initial plasma with the temperature $T=2.5$ and a field amplitude $E_0=0.5 E_{cr}$.  }
    \label{Fig6}
\end{figure}

We have seen in \cref{Fig3,Fig4} that pair creation gives a shorter period time (higher plasma frequency) and a reduced peak momentum compared to the classical predictions. Potentially, these deviations might be underestimated when using PIC codes or other semi-classical approaches to simulate plasma dynamics subjected to fields close to the critical field. To better understand how much the predictions from the quantum model deviate from the classical one, and how this scale with the field amplitude and the plasma density, we define the deviation in period time $\delta T$. 
\begin{align}
    \delta T&=\frac{T_{q}-T_c}{T_q}
\end{align}
where $T_q$ and $T_c$ are the period time for the quantum and classical models, respectively.
Similarly, we define the deviation in the peak average (over the particles) gamma factor between the two models according to 
\begin{align}
    \delta \gamma&=\frac{<\gamma_q>-<\gamma_c> }{<\gamma_q>}
\end{align}
We note that for $\max A\gg1$ (which is the regime of interest), the peak average gamma factors coincide with the values of $A$ when $E=0$.

In \cref{Fig5}, we plot the relative error in period time $\delta T$ and average gamma factor $\delta \gamma$ as a function of the initial field amplitude. Note that the values of $\delta T$ have been based on the first two peaks of a plasma oscillation, and the value of  $\delta \gamma$ is based on data from the second peak. For a longer numerical run lasting multiple periods, naturally, the relative error of $\delta T$ and $\delta \gamma$ for later times would have larger values than the ones shown here, as the number of particles in the system increases gradually. 

We can see in \cref{Fig5} that the relative errors are higher for stronger fields, as expected. The error is just a few percentages for fields lower than $0.4 E_{cr}$. However, we have used a plasma with a number density $n=2\times 10^{29}/{\rm cm^3}$, much larger than solid density. For substantially lower plasma densities, the relative errors will be higher, as the error depends on particles created in relation to those already present.  To explore this further, we plot in \cref{Fig6}, the relative error $\delta T$ and $\delta \gamma$ as a function of the plasma density using $E=0.5 E_{cr}$. By decreasing the plasma density by a factor 5, the relative error becomes 10 times higher. For even lower plasma densities, one can suspect that the relative error would be large enough to make the classical model completely inapplicable.

To be able to identify how large relative errors $\delta T$ and $\delta \gamma$ are for lower plasma densities than we have used so far, we have to overcome the problem of the large cutoff parameters. For low plasma densities, even fields weaker than $0.1 E_{cr}$ result in too large gamma factors, such that numerical solutions for the DHW-equation become increasingly memory-demanding and start to be unstable before having enough data to analyze.
Specifically, when the gamma factor is larger than 200, the period time of the plasma oscillation is long enough that the numerical instability arises before reaching a quarter period, using time and momentum steps that can be run on a PC. For a plasma density $n=10^{27}/{\rm cm^3}$ or lower and a field amplitude that is stronger than $0.1 E_{cr}$, the gamma factor is 1000 or higher. To study the Schwinger pair creation at this low-density plasma regime, we will make use of a simplified hybrid model. The idea is to use the classical Vlasov equation to determine the electric field evolution and to predict the pair creation rate $n$ based on the standard expression
\begin{equation}
   \frac{dn}{dt}= E^2 e^{-\pi/E}
   \label{rate-eq}
\end{equation}
We will refer to this hybrid model as the VPP (Vlasov-Pair-Production) approximation. While a rigorous derivation leading to  \cref{rate-eq} (see e.g. \cite{nikishov1970pair}) can be made, the production rate does not account for the complete dynamics. In particular, the hybrid model will just predict the production rate without accounting for the dynamical contribution from the newly created pairs. By necessity, however, this effect is a minor perturbation as long as the number of newly created pairs is small compared to those present initially. However, provided the initial number density is not small compared to the degeneracy level, the effect of Pauli blocking (included in the DHW theory) might still be substantial. Thus, before drawing conclusions based on the VPP approximation, we should first validate the production rate in the hybrid model by comparing generated pairs with that seen in the full DHW theory.  

\begin{figure}
    \centering
    \includegraphics[width=9cm, height=9cm]{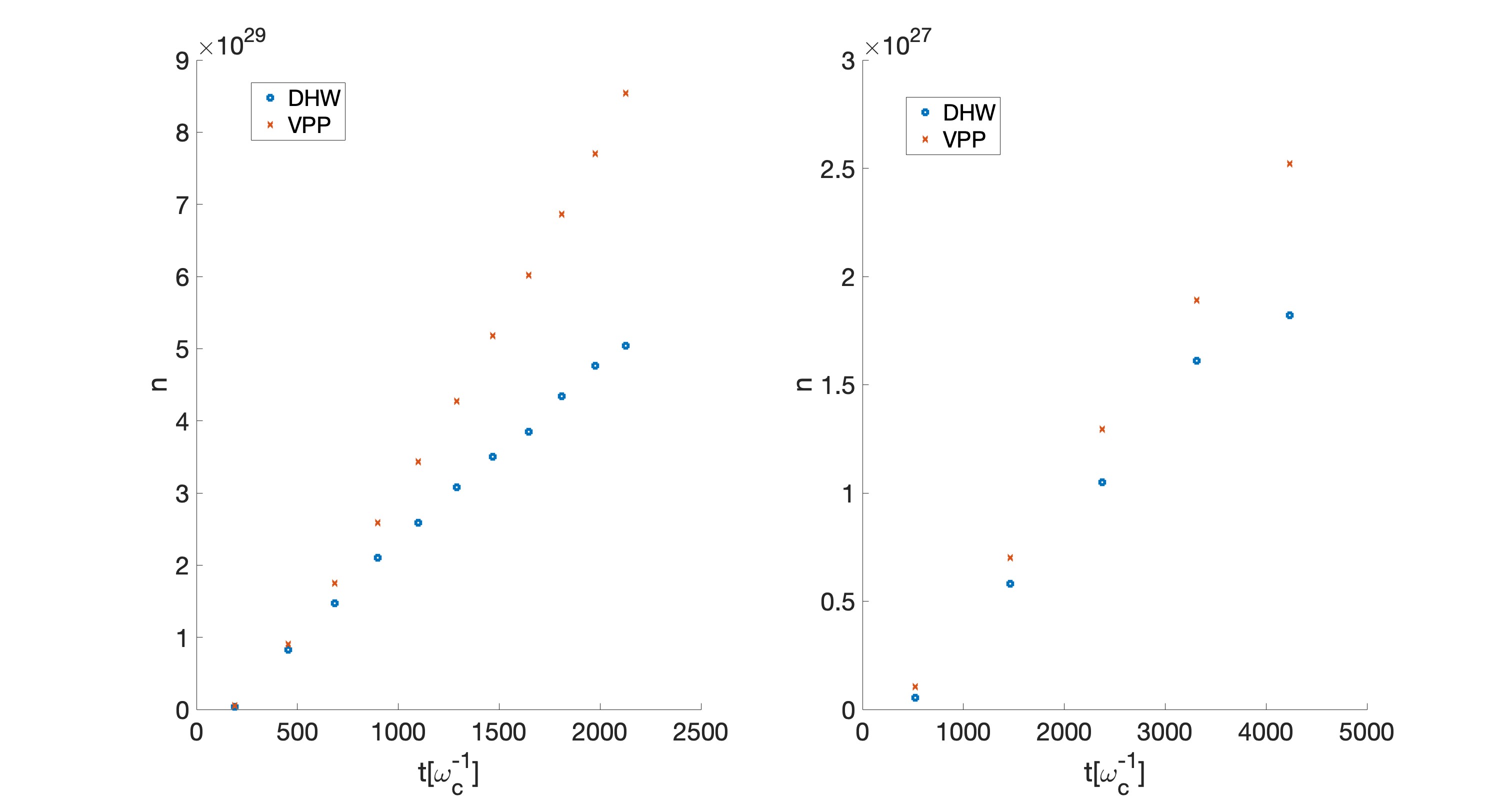}
    \caption{The number of produced particles $n$ for DHW and VPP using two different field amplitudes. For the left panel, we used $E=0.38 E_{cr}$ and for the right panel, we had $E=0.8 E_{cr}$.  }
    \label{Fig7}
\end{figure}

For this purpose, we compare the pair creation using both the DHW-equation and the VPP approximation, starting with the case of  $E=0.82 E_{cr}$ and a plasma density $n=5.8\times 10^{29}/{\rm cm^3}$, as shown in left panel of \cref{Fig7}, corresponding to the same electric fields evolution as shown in the upper panel of \cref{Fig3}. The number of pairs created for the two models are evaluated a half-period apart, at the times where $E=0$ for the respective calculation. We note that the production rate is more or less constant, not only in the VPP approximation but also in the DHW model (as the electric field decays very slowly). However, in the left panel, the production rate is significantly larger in the VPP approximation. This should not be a surprise, as the initial number density is large enough to block a significant part of momentum space where pairs would otherwise be created, an effect that is taken into account in the DHW theory. Next, we turn our attention to the case of a lower electric field and, importantly, to a lower initial density,  $E=0.38 E_{cr}$ and $n=8\times 10^{28}/{\rm cm^3}$, as shown in the right panel of \cref{Fig7}. As expected, the difference in production rate is smaller this time, as the lower initial density does not block as much of momentum space. Note, however, that the improved agreement is not completely obvious, as a reduced electric field to some extent limits the region in momentum space where pairs can be created. Despite this potential caveat, the increased accuracy of the VPP approximation with gradually decreasing initial density can be confirmed.                
 For even lower plasma densities and weaker fields, the Pauli blocking will be completely insignificant as the number of produced pairs is not high enough to fill the low-energy states, and the initial particle number also is too small to contribute to a substantial Pauli-blocking.

\begin{figure}
    \centering
    \includegraphics[width=8cm, height=8cm]{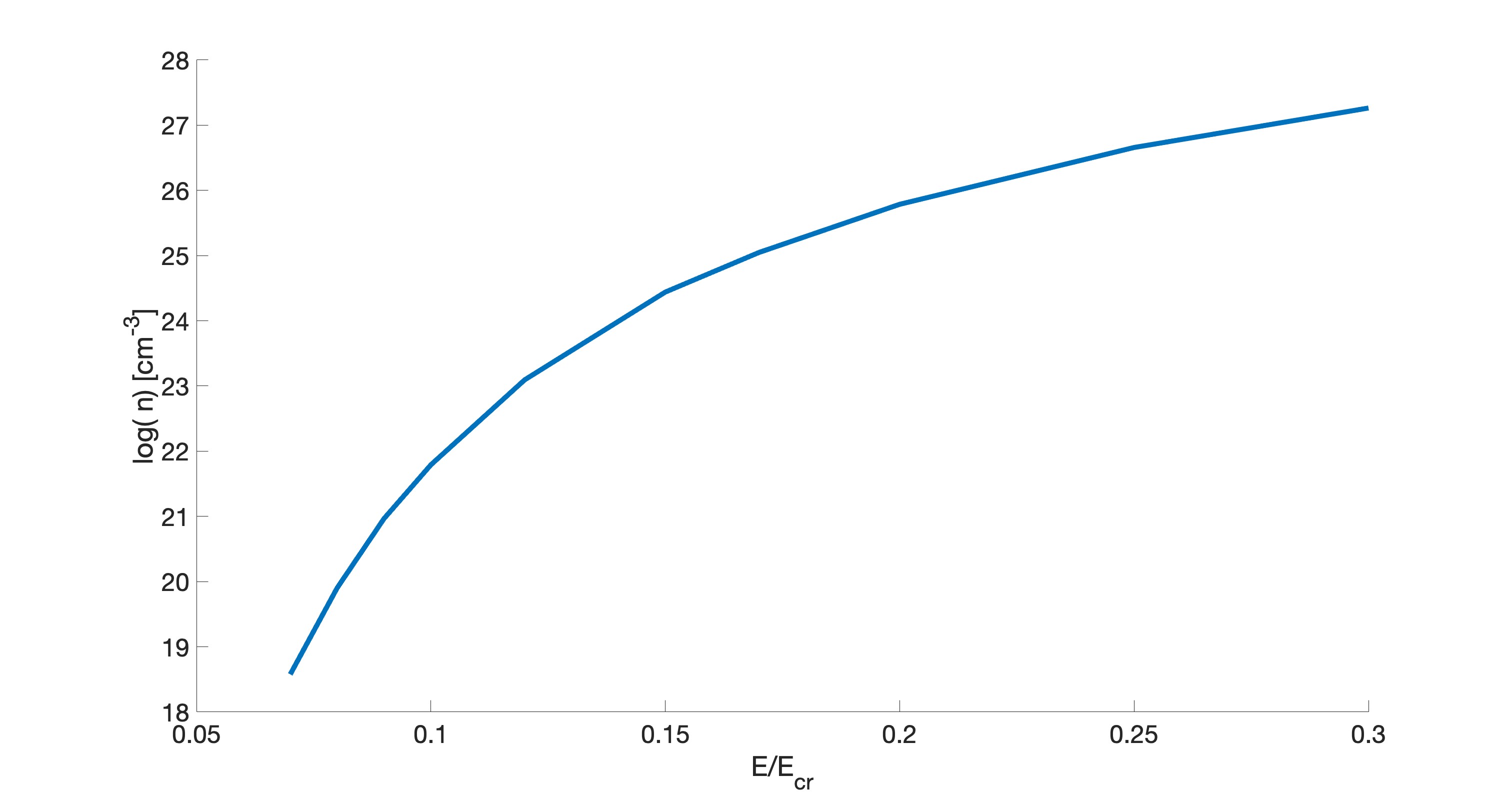}
    \caption{A demonstration of the breakdown of the Vlasov theory where we have defined that breakdown occurs when the VPP model predicts a 20\% increase in pairs per quarter period (that is, after a full period almost a doubling of particles). In the y-axis we have the plasma densities that result in breakdown as a function of the field amplitude shown on the x-axis. }
    \label{Fig8}
\end{figure}

Next, we can use the VPP approximation to predict the dynamics at even lower plasma densities, in order to investigate the eventual breakdown of the Vlasov theory.
The definition of "breakdown" is somewhat arbitrary. Even small deviations from the Vlasov theory that can result in a difference in the number density of a few percentages after one period will accumulate after many periods, due to sustained pair production. Eventually, this would result in completely incorrect values of the time period for the plasma oscillations. However, here we consider a more pronounced and dramatic breakdown, to be defined as follows: When the number of the produced pairs $n$ during 1/4 of the plasma period is $20\%$ of the initial plasma density $n_0$, we label this as a breakdown of the Vlasov theory, as the classical theory would not suffice even to get the first time period of the oscillation correct to a decent approximation. In \cref{Fig8}, we plot the plasma density vs. the field amplitude, where we would have a breakdown of the Vlasov theory according to the above definition. We notice that for a plasma density as low as $10^{18}/{\rm cm^3}$, the classical theory breaks down for a field of the order $0.07 E_{cr}$. Thus, using standard PIC-simulations for fields this large, only makes sense for number densities that are substantially higher. Similarly, a slightly larger field, around $0.1 E_{cr}$, will require a considerably larger number density, as we then would have a breakdown for $n\sim 10^{21}/{\rm cm^3}$.  

\section{Discussion}
\label{Discussion}
Quantum relativistic breakdown of classical evolution can occur for a multitude of reasons. In the present paper, we have considered an electrostatic geometry that suppresses the mechanisms of nonlinear Breit-Wheeler pair-production \cite{di2016nonlinear,seipt2020spin}, and of radiation reaction \cite{blackburn2020radiation,burton2014aspects}. A justification that such mechanisms are indeed small for the case of 1D plasma oscillations has been given in Ref. \cite{2023}. Another potential breakdown of the classical theory comes from spin-polarization, induced by strong fields \cite{crouseilles2023vlasov,del2017spin}, leading to polarization currents that in principle may compete with (classical) free current density. However, while spin polarization is a feature of the DHW theory, as long as we consider frequencies well below the Compton frequency and fields below the critical field, we have found that the polarization current of particles is negligible compared to the free currents. Moreover, under the same conditions, the free current of the DHW theory is well approximated by the classical Vlasov theory. In this context, one may note that the DHW-theory in addition to the polarization current of particles, also contains a vacuum polarization current. However, as long as we consider self-consistent plasma oscillations, it is worth noting that the vacuum contribution to the current density is small compared to the particle contribution, assuming that the Euler-Heisenberg Lagrangian (see e.g. \cite{marklund2006nonlinear}) gives the correct order of magnitude for the vacuum sources. 

From the above considerations, one might think that the classical approximation is valid for the electrostatic geometry, as soon as we have modest frequencies and fields well below the critical field. However, this is in fact not true. While the well-known exponential suppression of Schwinger pair production for fields below the critical field is indeed recovered in the numerical calculations, caution is still needed. The reason is simply that the pair-production rate for fields of the order of the Schwinger critical field is very high in relation to typical plasma densities. Thus, even if the process is exponentially suppressed for fields well below the critical field, the number of pairs created can still be dynamically significant if the initial value of the number density is not high. In particular, we have shown that in spite of the exponential suppression of the Schwinger mechanism, the classical Vlasov theory breaks down due to pair-production for relatively modest fields, of the order $0.05-0.1 E_{cr}$, if we have plasma densities of the order $n\sim 10^{19}-10^{21} {\rm cm}^{-3}$.   

As it is numerically challenging to solve the DHW-equation in the low-density regime, when the particle motion becomes sufficiently ultra-relativistic, the above conclusions regarding the Vlasov breakdown have been made based on a classical-quantum hybrid model. A future problem of much interest would be to overcome the numerical challenges imposed by low densities and study the truly ultra-relativistic ($\gamma>1000$) quantum regime without such simplifications.

\section{acknowledgement}
 H. A-N acknowledges support by the Knut and Alice Wallenberg Foundation.
 \bibliography{refs}

\end{document}